\documentclass[aps,prb,twocolumn,showpacs]{revtex4}
\usepackage[pdftex]{graphicx}
\usepackage{txfonts}
\usepackage{color}
\usepackage{float}
\usepackage{nicefrac}
\usepackage{multirow}
\usepackage{capt-of}
\newcommand{\eun}{EuNiGe$_3$}
\newcommand{\mub}{$\mu_{\rm B}$}

\begin{document}

\title{Exploring metamagnetism of single crystalline \eun\ by neutron scattering}
	
\author{X. Fabr\`eges$^1$, A. Gukasov$^1$, P. Bonville$^2$, A. Maurya$^3$, A. Thamizhavel$^3$ and S. K. Dhar$^3$}
\address{$^1$ Laboratoire L\'eon Brillouin, CEA, CNRS, Universit\'e Paris-Saclay, CEA-Saclay, 91191 Gif-sur-Yvette, France}
\address{$^2$ SPEC, CEA, CNRS, Universit\'e Paris-Saclay, CEA-Saclay, 91191 Gif-Sur-Yvette, France}
\address{$^3$ Department of Condensed Matter Physics and Materials Science, Tata Institute of Fundamental Research, Homi Bhabha Road, Colaba, Mumbai 400 005, India}	

\begin{abstract}
We present here a neutron diffraction study, both in zero field and as a function of magnetic field, of the magnetic structure of the tetragonal intermetallic \eun\ on a single crystalline sample. This material is known to undergo a cascade of transitions, first at 13.2\,K towards an incommensurate modulated magnetic structure, then at 10.5\,K to an equal moment, yet undetermined, antiferromagnetic structure. We show here that the low temperature phase presents a spiral moment arrangement with wave-vector {\bf k} = ($\frac{1}{4},\delta,0)$. For a magnetic field applied along the tetragonal {\bf c}-axis, the square root of the scattering intensity of a chosen reflection matches very well the complex metamagnetic behavior of the magnetization along {\bf c} measured previously. For the magnetic field applied along the {\bf b}-axis, two magnetic transitions are observed below the transition to a fully polarized state.
\end{abstract}

\maketitle

\section{Introduction}
Neutron diffraction on Eu materials is inherently  difficult because of the very strong absorption cross section of natural europium. Nevertheless, magnetic structure determinations were carried out a few decades ago in single crystalline EuAs$_3$ \cite{chatto1,chatto2} and in EuCo$_2$P$_2$ \cite{reeh}. Interestingly, antiferromagnetic EuAs$_3$ presents a feature which was to be found in many Eu intermetallics studied later: a first transition to an incommensurate phase, extending only over a few K, followed by a transition to an equal moment phase \cite{eupdsb,euptsi3,euptge3,eutal4si2,eutsi3,euirge3}. But most of the information about the magnetic structure of Eu compounds has been quite often inferred only through single crystal magnetization measurements or M\"ossbauer spectroscopy on the isotope $^{151}$Eu, like in EuPdSb \cite{eupdsb}. In the last few years, however, neutron diffraction with thermal neutrons was successfully employed to unravel the magnetic structure of some intermetallic divalent Eu materials \cite{ryan1,ryan2,ryan3,ryan4}. Of the two valences Eu$^{3+}$ and Eu$^{2+}$, only the divalent, with a half-filled $4f$ shell with L=0 and S=7/2, has an intrinsic magnetic moment of 7\,\mub. Despite the quite weak anisotropy of the Eu$^{2+}$ ion due to its vanishing orbital moment, a variety of structures was found, ranging from ferromagnetic in EuFe$_2$P$_2$ \cite{ryan1} and Eu$_4$PdMg \cite{ryan3}, collinear antiferromagnetic (AF) in EuCu$_2$Sb$_2$ \cite{ryan4} to incommensurate spiral in EuCo$_2$P$_2$ \cite{reeh} and EuCu$_2$Ge$_2$ \cite{ryan2}. This indicates that the interionic interactions are quite complex in Eu intermetallics, most probably due to the oscillating character of the RKKY  exchange and also to the relative importance of the dipole-dipole interactions between rather large Eu$^{2+}$ moments of 7\,\mub. As a result, the deduction of their magnetic structure from solely macroscopic measurements is often impossible.\\
	
\begin{figure}
	\begin{center}
	\includegraphics[width=0.40\textwidth]{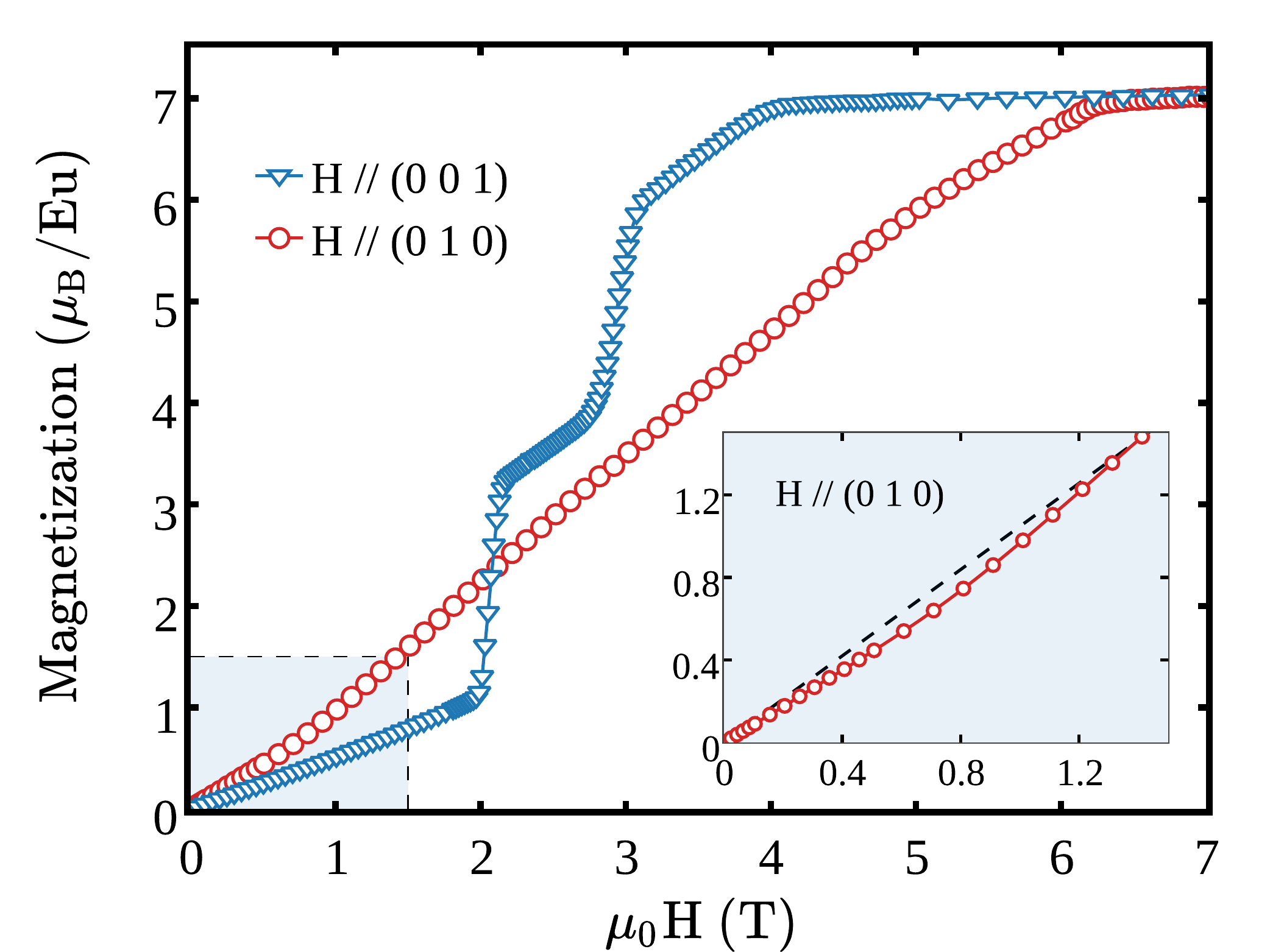}
\caption{\label{SQUID}Magnetization curves at 1.8\,K in \eun\ with the field along {\bf b} and {\bf c} taken from Ref.\onlinecite{maurya}. The insert shows the low field part of the curve for {\bf H} // {\bf b}, where a dip is clearly seen.}
	\end{center}
\end{figure}

In this work, we present a neutron diffraction study of single crystalline \eun. \eun\ was the subject of two previous studies, on a polycrystalline sample \cite{goetsch} and on a single crystal \cite{maurya}. It crystallizes in a body-centered tetragonal structure (space group $I4mm$) and presents two magnetic transitions, at $T_{N1}$=13.2\,K from the paramagnetic phase to an incommensurate moment modulated phase, then at $T_{N2}$=10.5\,K to an equal moment antiferromagnetic (AF) phase. The single crystal magnetization curve with field applied along the tetragonal {\bf c} axis shows a particularly complex behavior at 1.8\,K \cite{maurya}, with two spin-flop like magnetization jumps at 2 and 3\,T followed by a saturation in the field induced ferromagnetic  phase at 4\,T (see Fig.\ref{SQUID}). When the field is applied along the {\bf a} ({\bf b}) axis, the magnetization curve shows no such anomaly and reaches saturation at 6\,T. However, a small deviation from linearity is observed for this direction at low field, as shown in the insert of Fig.\ref{SQUID}, and the linear behaviour is recovered above 1.3\,T.

Assuming simple AF structures with propagation vectors {\bf k} = (001) or ($\nicefrac{1}{2}\ \nicefrac{1}{2}\ 0)$, it was not possible to reproduce the magnetization curve along {\bf c} using a molecular field model involving two nearest neighbor exchange constants (along {\bf a} and along {\bf c}), the dipolar field and a weak crystal field interaction \cite{maurya}. Clearly, an experimental determination of the zero field magnetic structure is needed in order to go further in the understanding of \eun. This was the original aim of the present work, but while exploring the in-field metamagnetic behavior of \eun, we have found a number of magnetic phase transitions which were not detected by  macroscopic measurements. Here we give the detailed description of these transitions with the field oriented along the {\bf b} ({\bf a}) and {\bf c} directions and present a molecular field model with 4 exchange integrals which partially succeeds in reproducing both the zero field structure and the magnetization curves.
	 
\section{Experimental details}
	
Details of the preparation method of the \eun\ single crystals, grown with In flux, can be found in Ref.\onlinecite{maurya}. For the neutron diffraction study, a 3x3.7x1\,mm$^{3}$ single crystal was mounted with the {\bf c} axis or the {\bf b} axis vertical in the variable temperature insert of a 7.5\,T split-coil cryomagnet. Experiments were performed on the neutron diffractometer Super-6T2 (Orph\'ee-LLB) \cite{Super6T2}. Scattering intensity maps were measured at $\lambda$ = 0.902\,\AA\ (Cu monochromator and Er filter) by rotating the sample around the vertical axis with 0.1$^{\circ}$ steps and recording the diffraction patterns with a position-sensitive detector (PSD). This allowed to detect all transformations of the magnetic structure under magnetic field by direct inspection of the 3D crystal reciprocal space obtained by transformation of the measured sets of PSD images. For quantitative refinements and studies of the magnetic field dependence, single (lifting) counter mode was used. The results were analyzed using the Cambridge Crystallography Subroutine Library(CCSL) \cite{ccsl}.\\
	
Prior to magnetic structure studies, the nuclear structure was verified in zero field at 15\,K. A total of 213 reflections were measured and 94 unique ones (74 $>$ 3$\sigma$) were obtained by merging equivalents, using space group $I4mm$. Since Eu is a strongly absorbing neutron material, the absorption corrections are of major importance in the merging procedure.They were made using the ABSMSF program of the CCSL which properly accounts for the crystal shape. The absorption correction was found very important, yielding an absorption coefficient $\mu$=1.05\,mm$^{-1}$, which resulted in up to 80\% reduction in the intensity of some measured reflections. Absorption correction yielded an improvement of the internal factors of nuclear reflections from $R_{int}=0.32$ (without corrections) to $R_{int}=0.07$ and it was applied to all measured magnetic data sets. The nuclear structure parameters obtained in the refinement were found in agreement with those published earlier \cite{maurya}, with lattice parameters $a=b=4.34(5)$\,\AA\ and $c=9.90(5)$\,\AA. Extinction corrections were applied using the EXTCAL program of the CCSL, which takes into account the crystal shape. The extinction parameters and the scale factor obtained in the refinement of the nuclear structure were used as input in further magnetic structure refinements. 

\section{The magnetic structure in zero field}
	
\begin{figure}
	\begin{center}
	\includegraphics[width=0.45\textwidth]{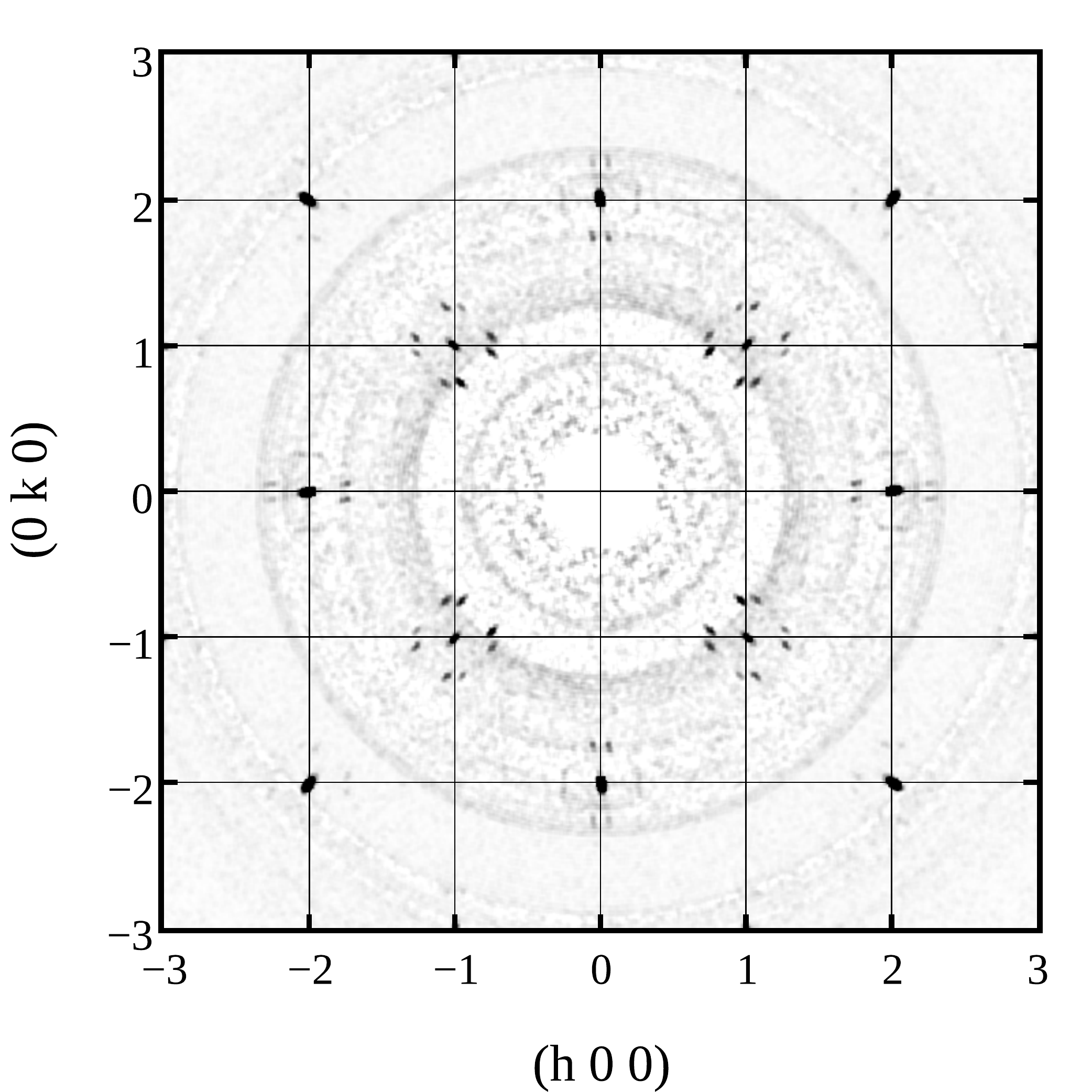}
\caption{\label{Tdep_slice}Nuclear reflections and magnetic satellites at 1.6\,K in \eun\ in the (h k 0) plane. Satellites observed around (1 1 0) can be indexed with a {\bf k} = (\nicefrac{1}{4} $\delta$ 0) propagation vector.}
	\end{center}
\end{figure}

The zero field magnetic structure of \eun\ was first studied using a PSD. Figure \ref{Tdep_slice} shows a bidimensional (h k 0) intensity cut in the reciprocal space at 1.6\,K. Apart from the nuclear reflections being located at integer positions, there are additional satellites which can be assigned to an antiferromagnetic contribution. Eight magnetic satellites can be distinguished around (1~1~0) and indexed using a {\bf k}=($\pm$\nicefrac{1}{4} $\pm\delta$ 0) propagation vector, with $\delta$=0.05, and its tetragonal permutations. In the following, the four possible {\bf k} domains are labelled {\bf k}$_1$=$\pm$(\nicefrac{1}{4} $\delta$ 0), {\bf k}$_2$=$\pm$(\nicefrac{1}{4} -$\delta$ 0), {\bf k}$_3$=$\pm$($\delta$ \nicefrac{1}{4} 0) and {\bf k}$_4$=$\pm$(-$\delta$ \nicefrac{1}{4} 0). These satellites form a star of the $I4mm$ space group and correspond to a rather complex antiferromagnetic structure with a very large unit cell, whose details are discussed below. For instance, {\bf k$_1$} = (\nicefrac{1}{4}~$\delta$~0) corresponds to a magnetic cell four times larger than the crystallographic one along {\bf a} and 20 times along {\bf b}. Actually, it is not possible to decide whether {\bf k} is incommensurate with the lattice spacing or not although, generally, such a small $\delta$ value points to an incommensurate structure.

\begin{figure}
	\begin{center}
	\includegraphics[width=0.35\textwidth]{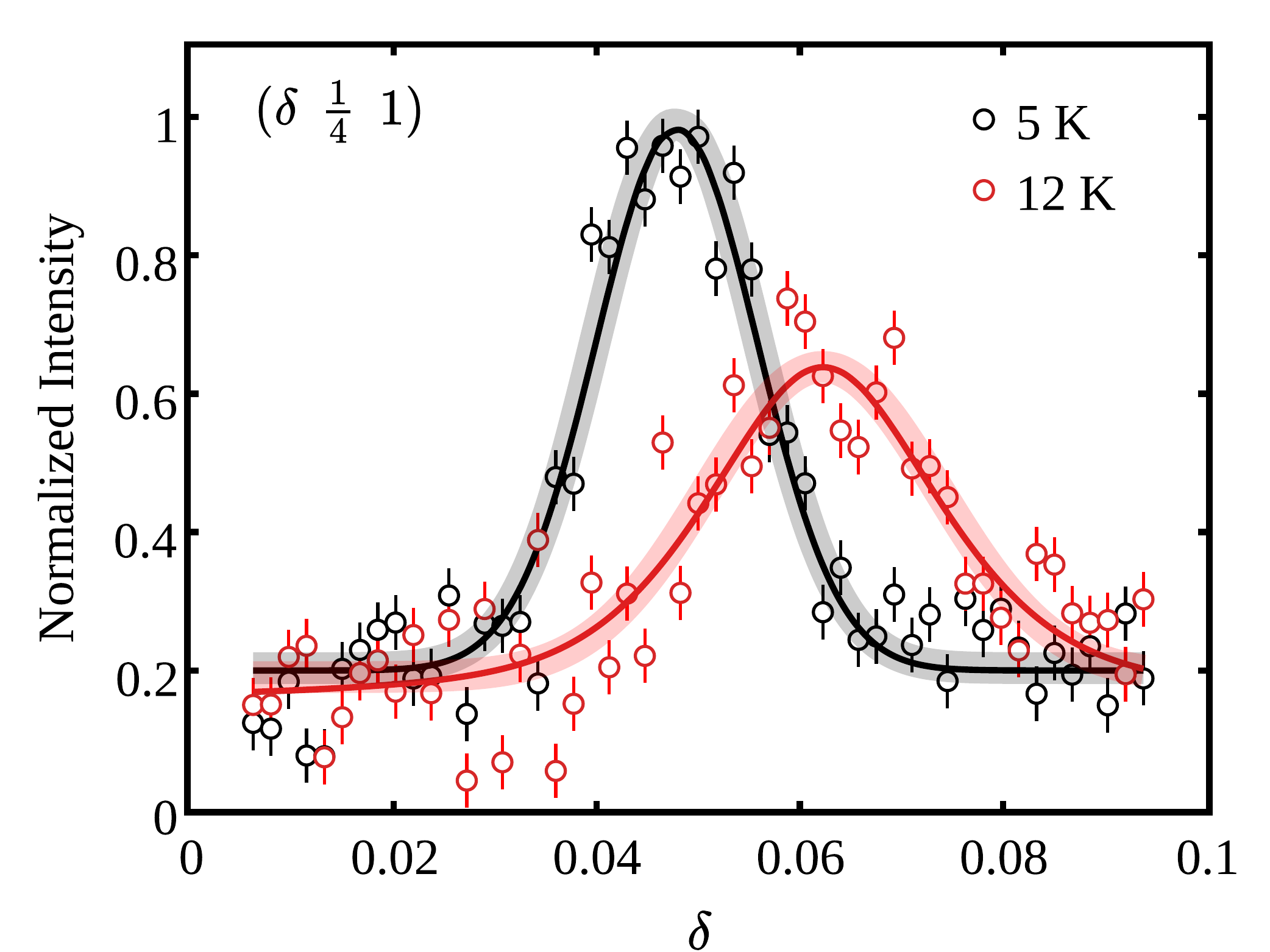}
\caption{\label{T_inc} Position of the magnetic satellite at 5 and 12\,K. At 12\,K a clear shift of the $\delta$ value is observed.}
	\end{center}
\end{figure}

The temperature evolution of {\bf k$_3$}=($\delta$~$\nicefrac{1}{4}$~0) magnetic reflections were followed to monitor the transitions from the antiferromagnetic to the paramagnetic state. The value $\delta \simeq$0.05 remains unchanged up to about 11\,K, and undergoes a small shift to 0.066 at 12\,K, as clearly seen on Fig.\ref{T_inc}. Figure \ref{slice_delta} top shows that, up to 12\,K, the thermal variation of the scattering intensity can be well fitted to the S=7/2 mean field function adequate for Eu$^{2+}$ ions, with a molecular field constant $\vert \lambda \vert \simeq$5.95\,T/\mub. Such a fit gives an excellent agreement between calculated and experimental data, with a transition temperature T$_t \simeq$12.0\,K. Above 11\,K, the (1+$\delta$~$\nicefrac{1}{4}$~1) satellite intensity deviates from the mean field function, and vanishes above 13.5\,K. In this temperature range, the observed small shift of the $\delta$ value corresponds to the intermediate phase reported in the M\"ossbauer investigation \cite{maurya}, which shows an incommensurate modulation of Eu moments. Therefore, the value $\delta'$=0.066 does correspond to an incommensurate modulation, but the weakness of the magnetic signal in this phase prevented us from determining its detailed structure.	
	
\section{The field variation of magnetic structure}

\begin{figure}
	\center{
	\includegraphics[width=0.35\textwidth]{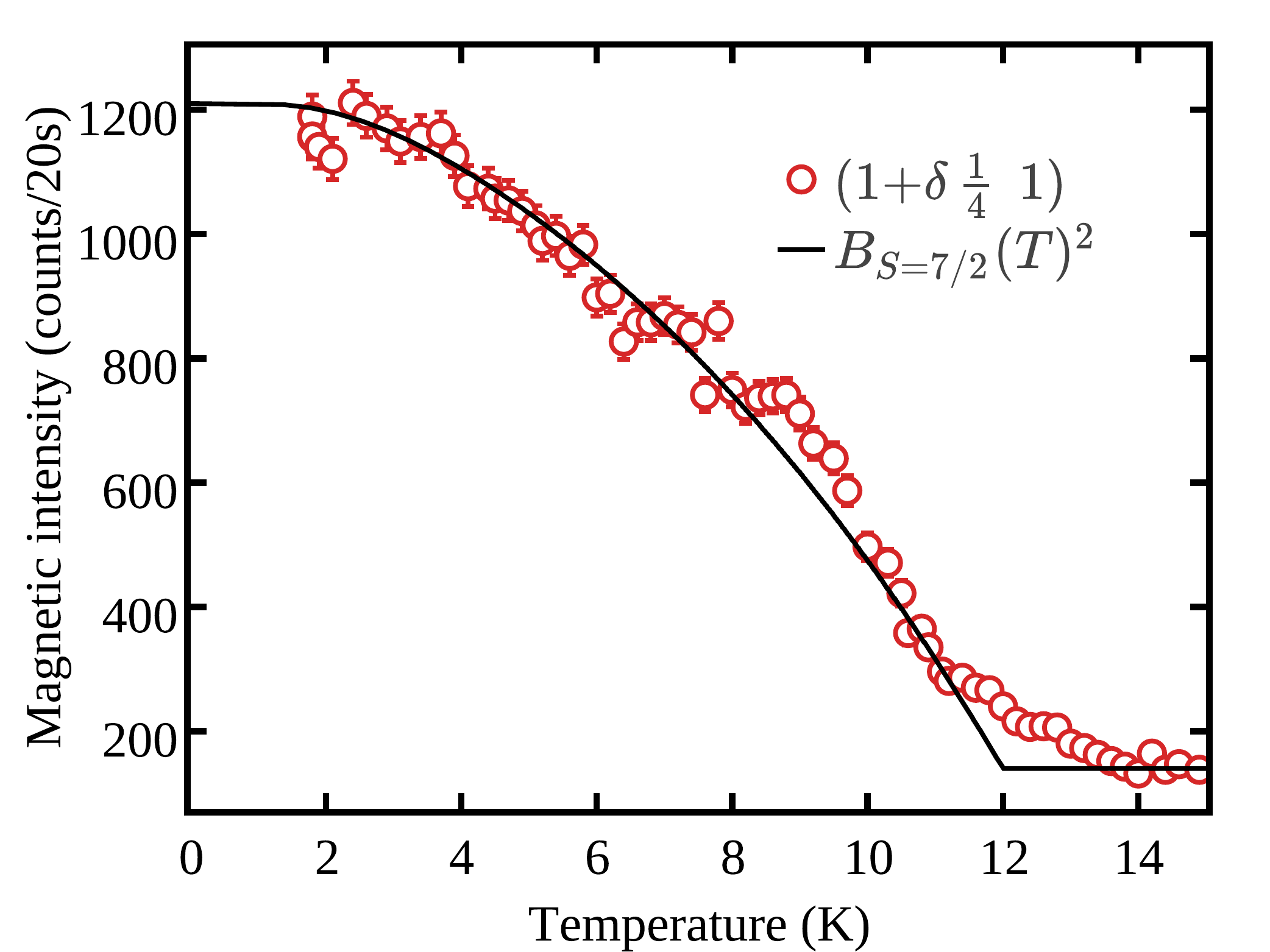}
	\includegraphics[width=0.35\textwidth]{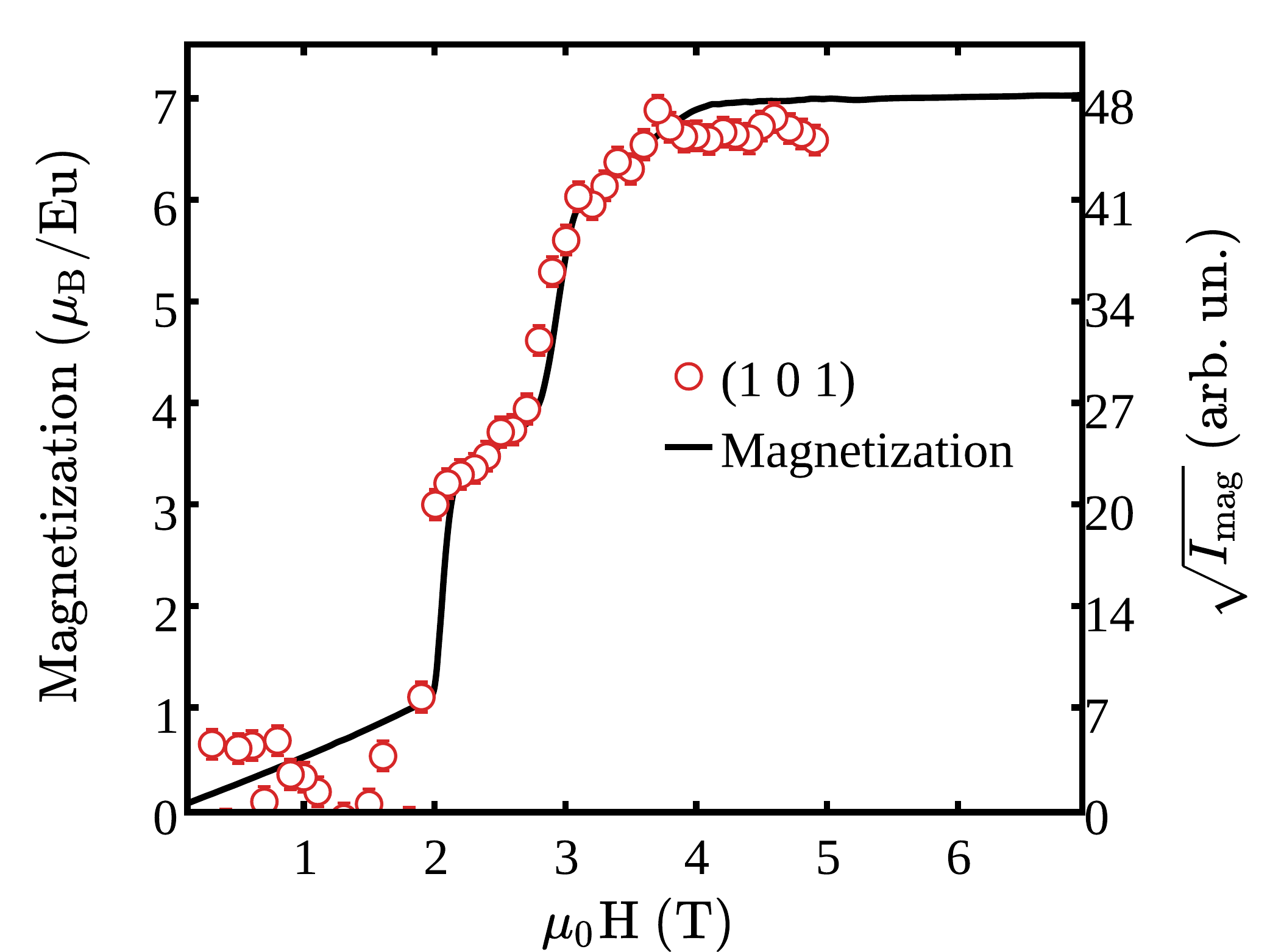}
\caption{\label{slice_delta}{\bf Top}: (1+$\delta$ \nicefrac{1}{4} 1) scattering intensity {\it vs} temperature (red circles) and fit to a (squared) S=7/2 mean field law (black line); {\bf Bottom}: magnetization at 1.8\,K (black line) and square root of the (1~0~1) scattered intensity at 1.6\,K (red circles) {\it vs} field applied along {\bf c}.}
		}
\end{figure}

\begin{figure}
	\center{
	\includegraphics[width=0.35\textwidth]{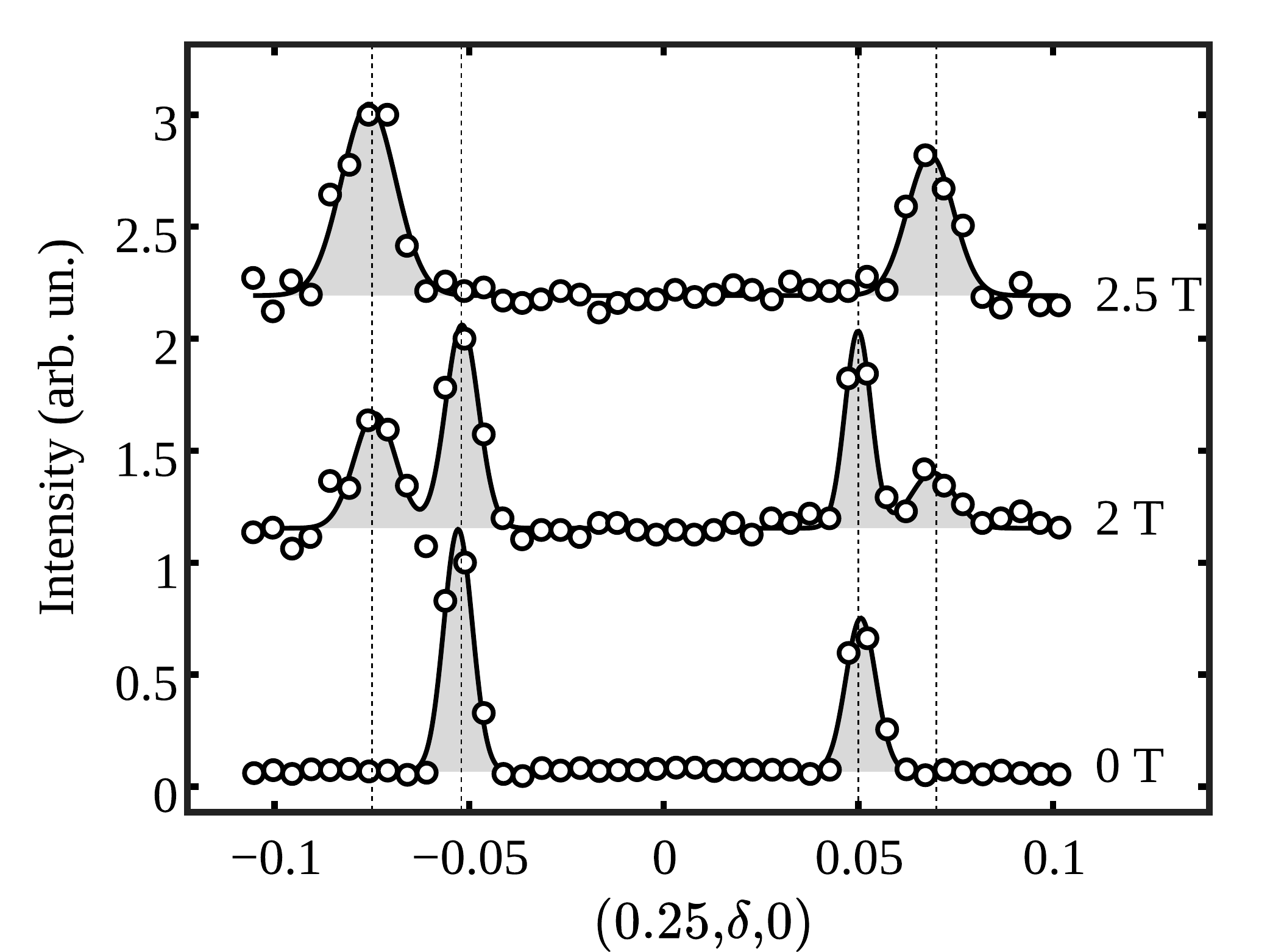}
	\includegraphics[width=0.35\textwidth]{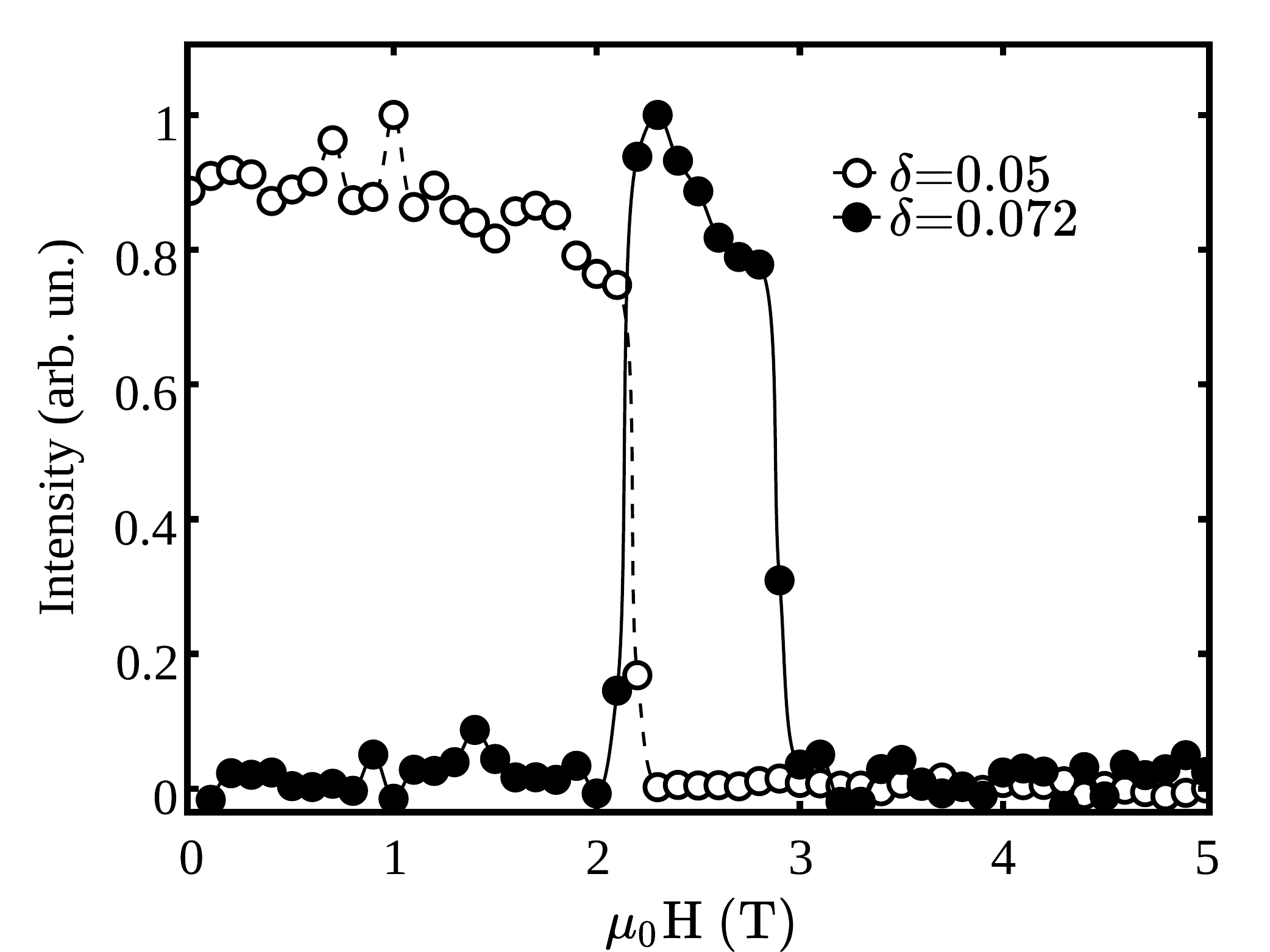}
\caption{\label{sweep}At 1.6\,K for {\bf H} // {\bf c}: {\bf Top}: (\nicefrac{1}{4}~$\delta$~0) scans at 0, 2 and 2.5\,T; {\bf Bottom}: (\nicefrac{1}{4}~$\delta$~0) scattering intensities {\it vs} field for $\delta$=0.05 (open circles) and $\delta$=0.072 (closed circles).}
		}
\end{figure}

With the magnetic field applied along {\bf c}, the behaviour of the magnetization is quite peculiar (see Fig.\ref{SQUID}). We monitored the scattering intensity of the (1~0~1) reflection for {\bf H} // {\bf c} as a function of field. This reflection contains both nuclear and magnetic contributions, the magnetic one being proportional to the square of the induced magnetization. Figure \ref{slice_delta} bottom shows the field evolution of the square root of the (1~0~1) magnetic scattered intensity (after substraction of the nuclear component) compared with the magnetization data. A very good agreement between the two probes is observed, with two well defined jumps at respectively 2 and 3\,T, followed by the spin-flip transition at H$\simeq$ 4\,T with the fully saturated Eu$^{2+}$ moment of 7\,\mub.	
The top panel of figure \ref{sweep} shows the scattering intensity along the (\nicefrac{1}{4}~$\delta$~0) direction at 1.6\,K for H=0, 2 and 2.5\,T. In zero field, two well defined peaks are observed at $\delta$=$\pm$0.05 confirming the splitting along {\bf b}$^*$. At 2\,T, the field of the first magnetization jump, a first order transition occurs with the appearance of two new satellites with $\delta^*$=0.072 coexisting with those at $\delta$=0.05. The new satellites correspond to a smaller magnetic unit cell in the {\bf b}$^*$ direction. At 2.5\,T, the zero field $\delta$=0.05 satellites  completely vanish. In turn, the $\delta^*$=0.072 satellites disappear at H=3\,T, corresponding to the field of the second magnetization jump. 

This evolution with the field is best evidenced by plotting the intensity of the two magnetic reflections corresponding to the propagation vector (\nicefrac{1}{4}~$\delta$~0), with $\delta$=0.05 and $\delta^*$=0.072 (fig.\ref{sweep} bottom). The intensity of the $\delta$=0.05 reflection disappears above 2\,T, while that with $\delta^*$=0.072 shows up. This one in turn vanishes at the second critical field of 3\,T, above which  only ferromagnetic (FM) contributions remain. Finally, the intensity of (FM) reflections reach saturation at 4\,T corresponding to the spin-flip field.

\begin{figure}
	\center{
	\includegraphics[width=0.35\textwidth]{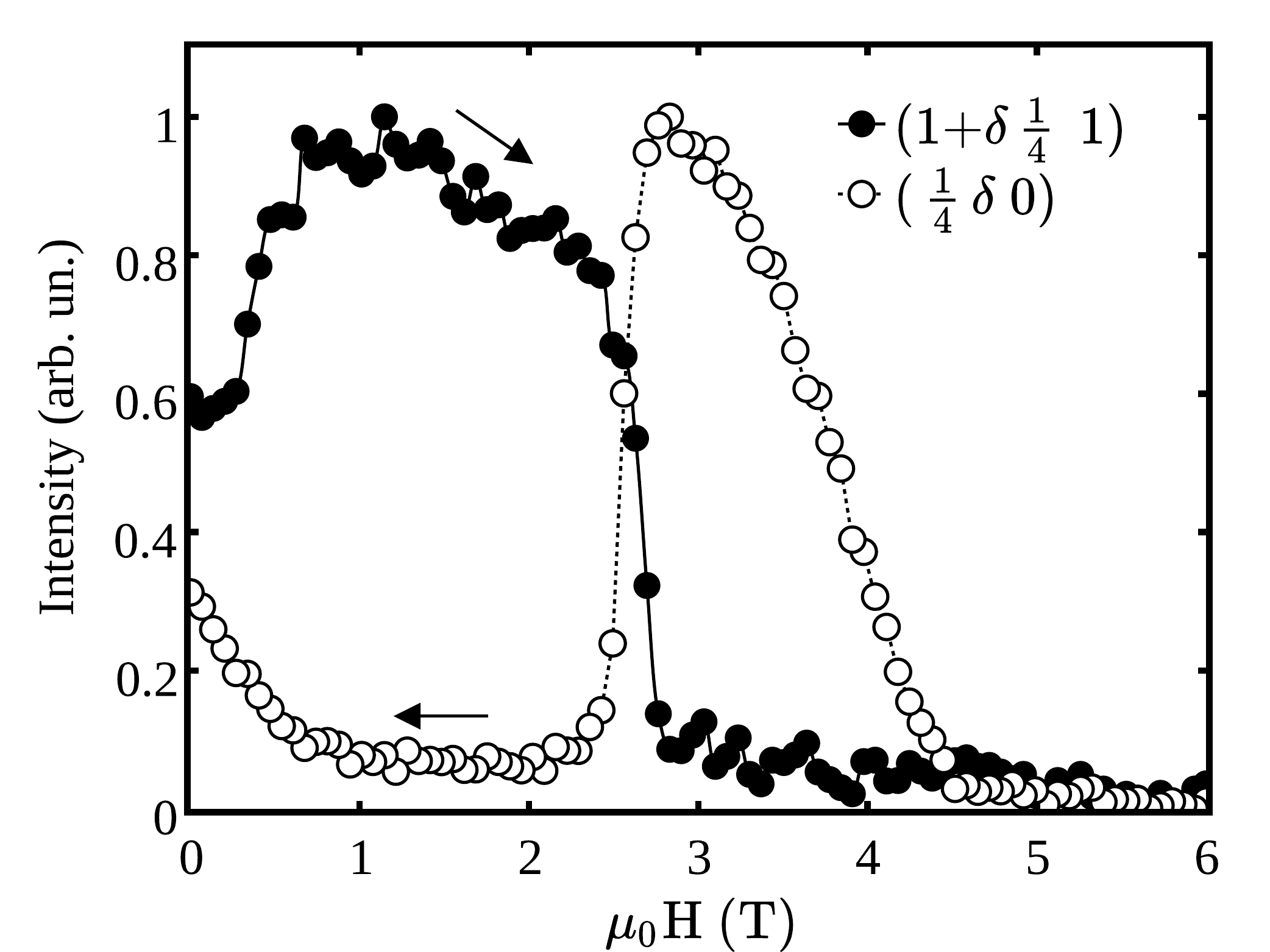}
	\includegraphics[width=0.35\textwidth]{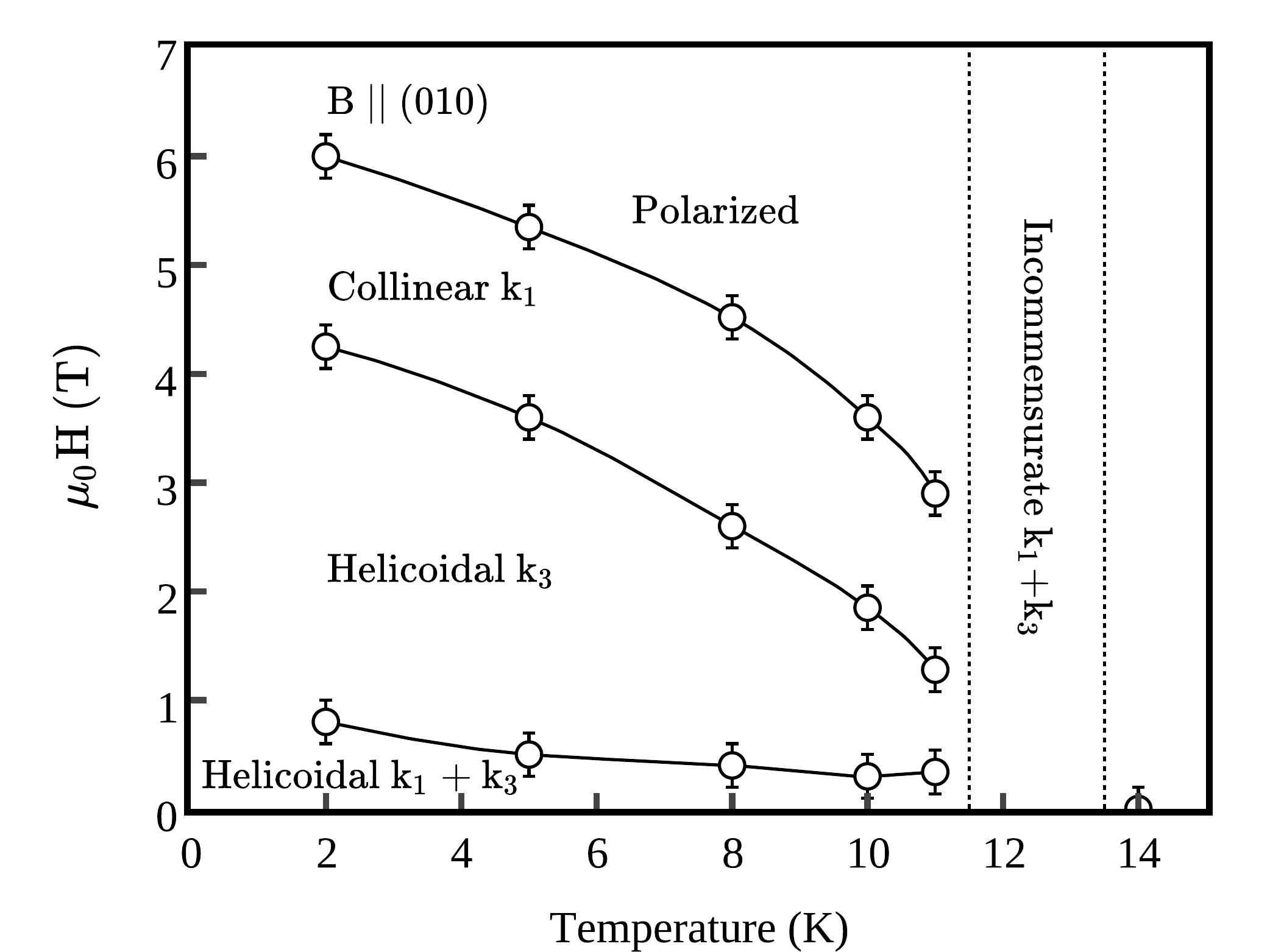}
\caption{\label{sat_B010}For {\bf H} // {\bf b}: {\bf Top}: Evolution of  k$_1$ and k$_3$ satellites versus magnetic field at 8\,K. Intensities have been normalized to one in both cases for clarity. Three transitions are observed at 0.4, 2.5 and 4.3T; {\bf Bottom}: corresponding (H,T) phase diagram.}
		}
\end{figure}

With the magnetic field applied along {\bf b}, we monitored the scattering intensities corresponding to the {\bf k$_1$}=($\nicefrac{1}{4}$~$\delta$~0) and {\bf k$_3$} = ($\delta$~$\nicefrac{1}{4}$~0) domains between 2 and 14\,K in fields up to 6\,T. Figure \ref{sat_B010} top presents the evolution of {\bf k$_1$} and {\bf k$_3$} intensities at 8\,K. Below $\mu_0H$=0.4\,T both reflections are observed with similar intensities as expected from the tetragonal space group. Above 0.4\,T, the intensity of {\bf k$_1$} vanishes at the benefit of {\bf k$_3$}. This first transition corresponds to a spin-flop transition selecting the (a,c) magnetic domains in which moments are orthogonal to the applied magnetic field. Above 2.5\,T the reverse process occurs with the sudden extinction of the {\bf k$_3$} signal at the benefit of {\bf k$_1$}.  Finally, no antiferromagnetic contribution is observed above 4.3\,T, the sample being fully polarized. The corresponding phase diagram extracted from neutron diffraction data is presented in Fig.\ref{sat_B010} bottom. Comparing with the phase diagram for {\bf H} // [100] in Ref.\onlinecite{maurya}, extracted from macroscopic measurements, one sees that the latter could not catch the first transition at low field. Besides this, the overall agreement is good, except for the behaviour near the transition at $T_{N1} \simeq 10.5$\,K which is shifted by about 0.5\,K in the neutron data.

\section{Magnetic structures refinement  }

For the zero field and in-field magnetic structure determination, integrated intensity measurements were performed using a single counter. 

\begin{figure}
	\center{
	\includegraphics[width=0.45\textwidth]{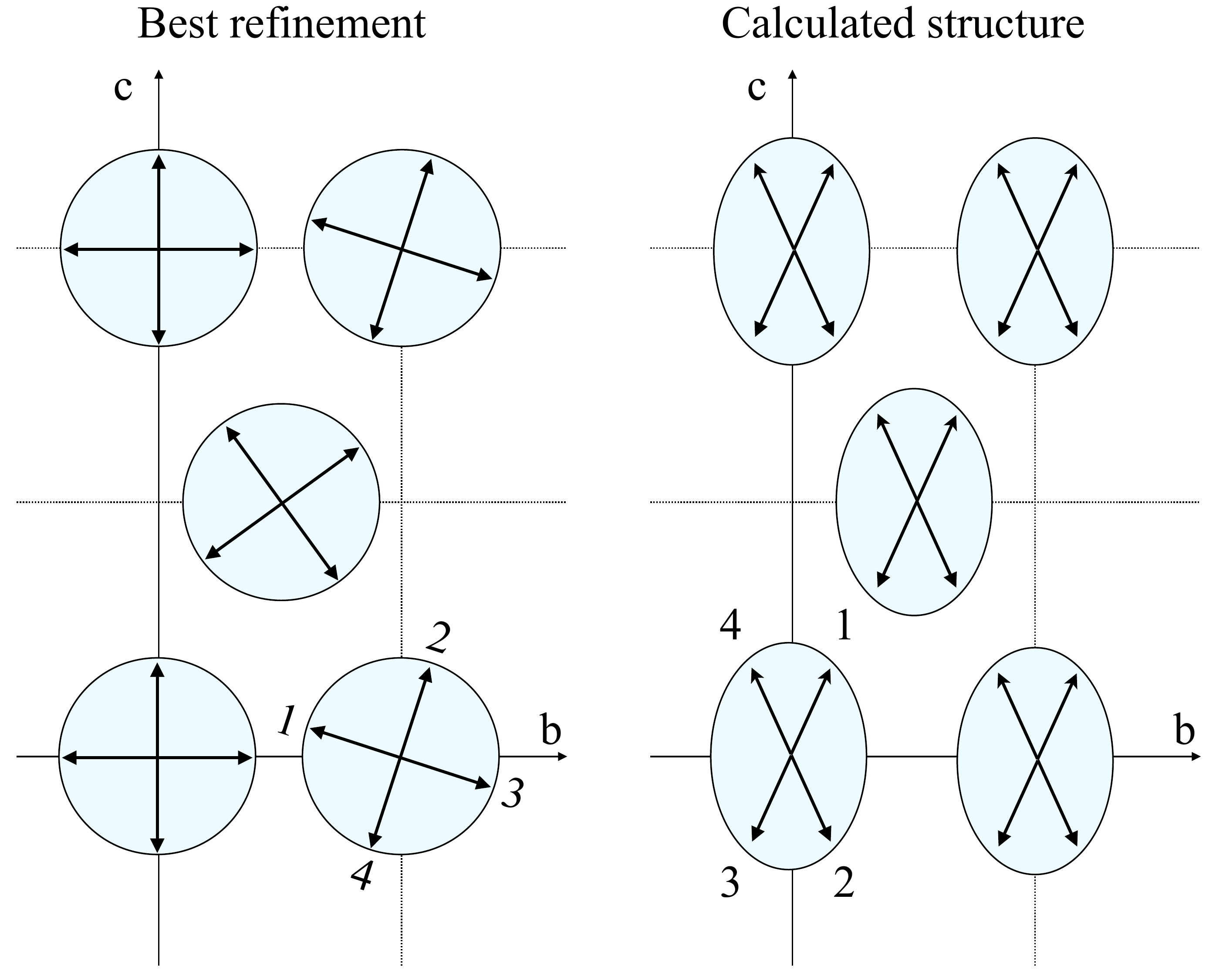}
\caption{\label{structa}a) {\bf k$_1$} refined magnetic structure in zero-field, with $\phi$=90$^\circ$. Note that other structures allowing for an ``elliptical'' envelope, i.e. with $\phi < 90^\circ$, are also possible. b) Calculated magnetic structure from the model with 4 exchange integrals, dipolar interactions and anisotropy described in section \ref{modl}. The labels 1-4 indices indicate the spin positions along the {\bf a}-axis, with $\phi$=65$^\circ$.}
		}
\end{figure}

In zero field, and for each {\bf k} domain, 46 satellites were collected at 1.6\,K, of which about 25 (0\,T)  were statistically relevant ($I>3\sigma$) and used in the refinement. The magnetic structure was analyzed by using the propagation vector formalism. Tetragonal symmetry and the highly symmetrical (0~0~z) Wykoff position occupied by the Eu$^{2+}$ ion limits possible magnetic structures to amplitude modulated and helicoidal ones. First, models of a circular helix with moments constrained in the planes perpendicular to the highest symmetry axes ({\bf a},{\bf b},{\bf c}) were tested. For all four  propagation vectors the best fit was obtained for the helix envelope with the major axes of 7.6(3)$\mu_B$ lying in the plane perpendicular to the largest component of the propagation vector, namely the (b,c) plane for {\bf k}$_1$ and {\bf k}$_2$, and the (a,c) plane for {\bf k}$_3$ and {\bf k}$_4$. In this case, the moment rotates by $\phi$=90$^\circ$ along the main component of the propagation vector. The corresponding magnetic structure is presented in Figure \ref{structa} left. The refinement yielded the following populations of domains in zero field: {\bf k}$_1$=35(5)\%, {\bf k}$_2$=30(1)\%, {\bf k}$_3$ =15(2)\%, and {\bf k}$_4$ =20(5)\% with 5.7\%$<R_w<$18.9\%. This is close to the expected random value of 25\% for all domains. Adding ``ellipticity'' to the helix (while maintaining a common modulus for the moment) yields a similar agreement factor. In this case, the angular increment along {\bf k} is alternatively $\phi$ and $\pi-\phi$. Thus, the incremental angle $\phi$ cannot be determined from the neutron data and both the circular and elliptic solutions are valid candidates.
	
With a field of 2.5\,T along the {\bf c}-axis (i.e. between the two metamagnetic transitions), for each {\bf k} vector, 46 satellites were collected at 1.6\,K, of which about twelve were statistically relevant and used in the refinement. In this case the antiferromagnetic contribution is well described  by a similar circular structure with reduced ordered magnetic moment of $m=5.5(5)$\,\mub. We note that the associated error bars are bigger, with $R_w=26.9\%$. This is due to a strong decrease of the antiferromagnetic signal after the first metamagnetic transition which resulted in the limited number of observed reflections. Finally, above 3\,T, a simple ferromagnetic contribution is observed reaching saturation at 4\,T.	
	
With the field along {\bf b}, integrated intensities of {\bf k$_1$} related reflections were collected at 5\,K and 4.5\,T. A total of 111 reflections were collected, of which 12 were statistically relevant ($I>3\sigma$) and used in the refinement. The antiferromagnetic contribution could not be refined with such a small set of reflections. However, neutrons are only sensitive to magnetic contributions orthogonal to the probed Q-vector. In our dataset, the ($\nicefrac{1}{4}$ $\delta$ L) reflections are not observed indicating the lack of an ordered magnetic moment in the orthogonal (a,b) plane. Therefore, one can describe the ordered moments as being antiferromagnetically coupled and collinear along the {\bf c}-axis. 
	
\section{Modelling the magnetic properties} \label{modl}

\subsection{The mean field self-consistent calculation}
	
We have searched for a set of exchange integrals that would reproduce the zero field magnetic structure and the behavior of the magnetization, using a self-consistent calculation of the moment arrangement in the presence of exchange and dipole-dipole interactions among Eu$^{2+}$ ions. Since this calculation cannot involve a large number of magnetic sites, it cannot integrate the $\delta$ components of the propagation vector evidenced by neutron diffraction in both zero-field and in-field structures. Based on the neutron diffraction results, we chose a propagation vector {\bf k} = ($\nicefrac{1}{4}$~0~0) and hence a magnetic lattice cell with dimensions (4a,a,c) containing 4 ions with $z$=0 and 4 ions with $z=\nicefrac{1}{2}$, i.e. 8 ions. The calculated structure consists of ferromagnetic (b,c) planes. The calculation does not either consider the twinning when computing the magnetization. Consequently, it cannot be expected to reproduce all of the experimental features and must be considered as approximate.

\begin{figure}
	\center{
	\includegraphics[width=0.4\textwidth]{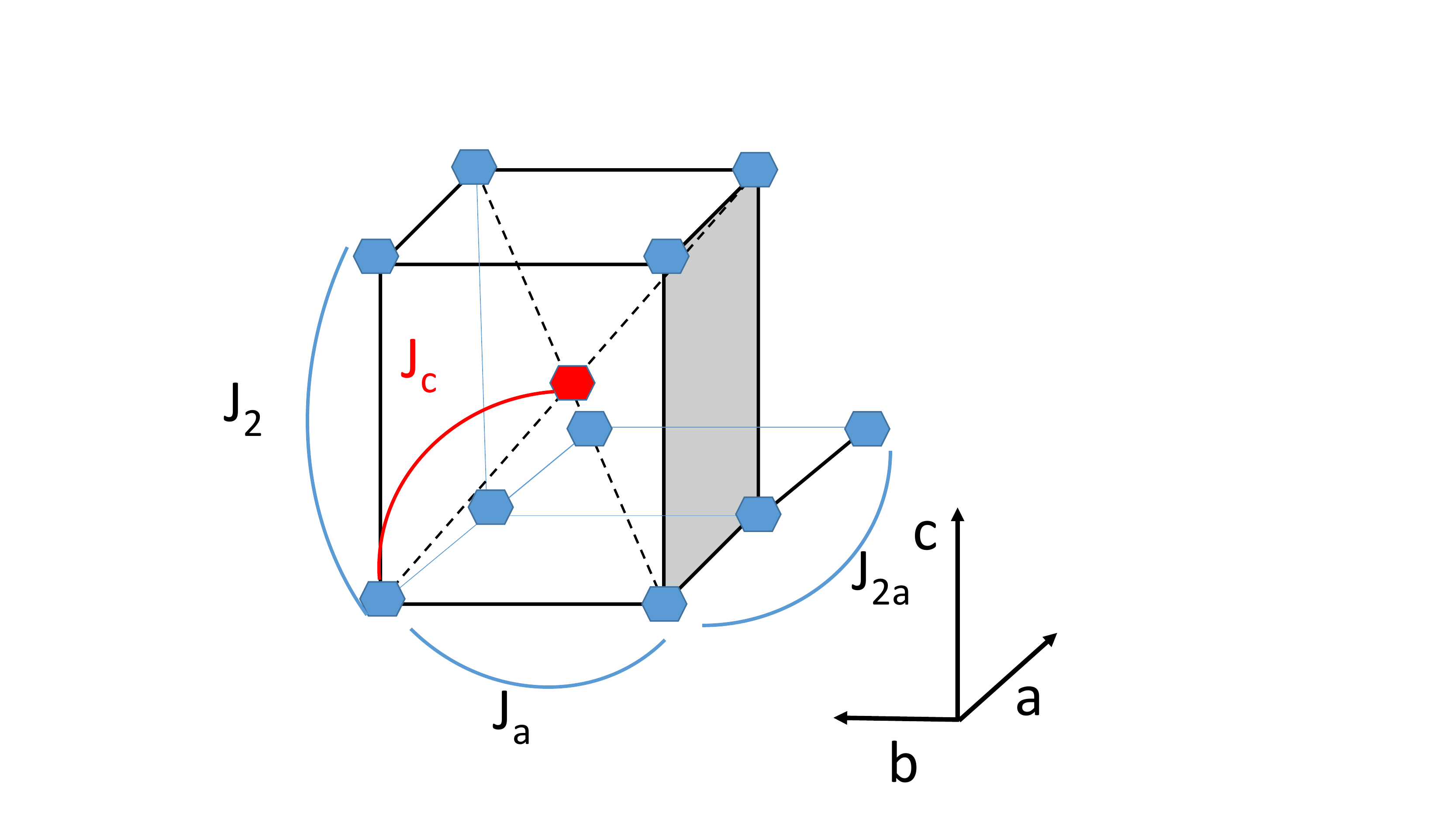}
\caption{\label{ech}Definitions of the 4 exchange integrals involved in the mean field calculations of the magnetic structure of \eun. The two simple tetragonal sub-lattices are sketched by blue and red hexagons.}
		} 
\end{figure}

We consider 4 isotropic exchange integrals, 3 of them intra-sublattice and one inter-sublattice, the sublattices in question being the simple tetragonal lattice and that obtained by the body centering translation with vector (\nicefrac{1}{2}~\nicefrac{1}{2}~\nicefrac{1}{2}) (see Fig.\ref{ech}).
The hamiltonian of the problem contains an exchange part (a negative integral means an antiferromagnetic coupling), a dipolar interaction part and also an anisotropy, or crystal field, part. The two latter terms are needed for a realistic description of the system since they are of the same order of magnitude and their balance determines the direction of the moments. Each ion is linked by exchange to its neighbors according to the paths described in Fig.\ref{ech}, and a molecular field is calculated for each of the 8 ions in the cell. The infinite range dipolar field acting on each ion is calculated using an Ewald type summation method \cite{wang}. The dipolar field has no free parameter, it depends only on the way the magnetic cell is chosen. An axial anisotropy (crystal field) term is added with the form: ${\cal H}_{an} = D S_z^2$, where $D$ is a coefficient with magnitude a few 0.1\,K and $S_z$ is the component along {\bf c} of the Eu$^{2+}$ spin. For $D < 0$, this term favors a moment alignment along {\bf c} and for $D > 0$ a moment arrangement in the (a,b) plane. The calculation is intended to reproduce not only the magnetic structure and the magnetization curves, with the correct spin-flip fields, but also the value of the N\'eel temperature $T_N \simeq$13\,K and of the paramagnetic Curie temperature $\theta_p \simeq$4\,K \cite{goetsch,maurya}. 

We have tried to obtain a zero-field structure like that shown in Fig.\ref{structa} a, which is the closest to the actual structure, neglecting the small $\delta$ component of the propagation vector. First, one finds that the $D$ coefficient must be taken negative, otherwise the moments have a strong affinity to lie in the (a,b) plane. Then, one must take $J_2 >$0 and $J_{2a} <$0 to obtain ferromagnetic (b,c) planes and alternating moment directions along {\bf a}. The other integrals $J_a$ and $J_c$ have no obviously required sign. Exploring the \{$J_\alpha$\} space of exchange integrals with reasonable values, we found that the magnetization jumps at 2 and 3\,T for {\bf H} // {\bf c} cannot be reproduced together. The parameter set we propose reproduces the spin-flop at an intermediate value 2.5\,T and the spin-flip at 4\,T for {\bf H} // {\bf c}, the spin-flip at 6\,T for {\bf H} // {\bf a} or {\bf b}, and the correct $T_N$ and $\theta_p$ values. It yields a zero-field structure of ``elliptic'' type, i.e. the moments lie in the (b,c) plane with an incremental angle $\phi$=65$^\circ$ both for the ions with $z$=0 and $z = (\nicefrac{1}{2}, \nicefrac{1}{2}, \nicefrac{1}{2}$) (see Fig.\ref{structa} b).
\begin{figure} [ht]
	\center{
	\includegraphics[width=0.45\textwidth]{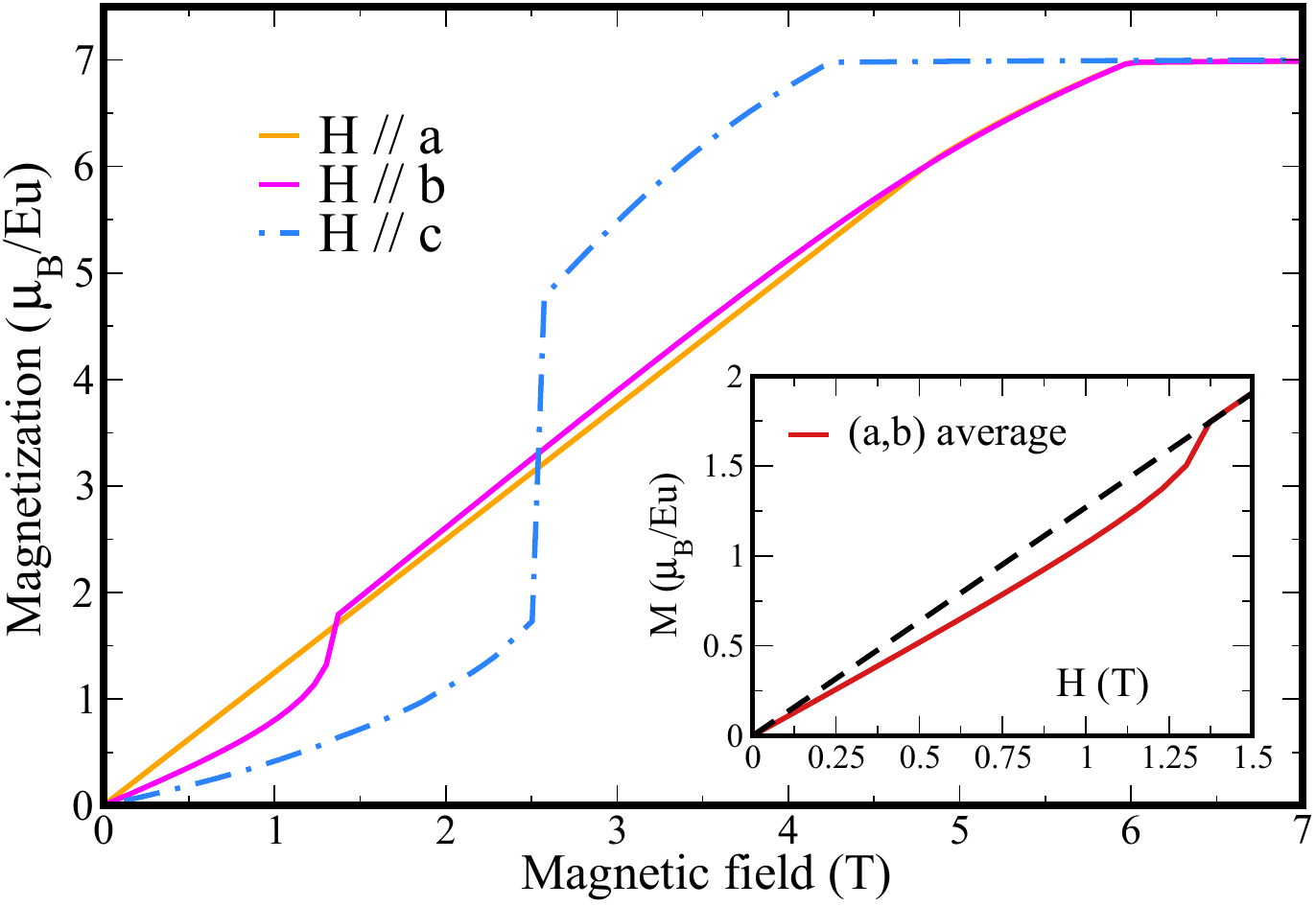}
\caption{\label{mh3dir} Calculated magnetization curves along the 3 symmetry directions for the spiral structure of \eun\ with {\bf k}=(\nicefrac{1}{4}~0~0), using the parameters given in the text. Insert: average of the magnetizations along {\bf a} and {\bf b} simulating the experimental data with {\bf H} // (100) or (010).}
		} 
\end{figure}
Neither the phase shift between the two spirals nor their absolute position can be ascertained in the calculation. The parameter set reads: $J_a$ = 0.1\,K, $J_c$ = 0.1,K, $J_2$ = 0.7\,K, $J_{2a}=-$0.6\,K and $D=-$0.25\,K. The calculated magnetization curves are represented in Fig.\ref{mh3dir}; they are to be compared with the experimental data in Fig.\ref{SQUID}. For {\bf H} // {\bf a} (orange curve), the magnetization is linear with the field, which is to be expected since {\bf H} is perpendicular to the (b,c) plane of the spiral, inducing a conical moment arrangement. For {\bf H} // {\bf b} (magenta curve), a dip is observed, which is due to the fact that the conical structure is not realized, when the field lies in the plane of the spiral, until a threshold spin-flop field is reached, here 1.3\,T. In the insert of Fig.\ref{mh3dir}, the red curve is an average of the magnetizations along {\bf a} and {\bf b}, simulating the presence of domains. It is readily comparable with the data shown in the insert of Fig.\ref{SQUID}, and the calculated spin-flop field of 1.3\,T is in very good agreement with the measured value. 

\subsection{Discussion}

In the above parameter set, the absolute value of $J_{2a}$, 0.6\,K, is six times larger than that of $J_a$, 0.1\,K. This may seem puzzling, since the next nearest neighbour distance along {\bf a} is twice the nearest neighbour distance. The dominant exchange in \eun\ is probably the RKKY interaction, which varies with distance as $1/r^3$, but which is an oscillating function of $r$. Then, one may speculate that a large $\vert J_{2a} \vert/J_a$ ratio can happen if RKKY exchange is close to a node for $r=a$ and is maximum for $r=2a$.

The Dzyaloshinski-Moriya (DM) exchange \cite{dm} was not included in the calculation, although the nearest neighbour ion pairs allow for a non-zero DM vector, their midpoint not being an inversion center. Introduction of DM exchange could induce the observed incommensurability, but it is likely that it cannot account for another puzzling feature of the magnetic structure of \eun: the symmetry breaking between the {\bf a} and {\bf b} axes. Indeed, the zero-field propagation vector, for instance {\bf k}$_1$ = (\nicefrac{1}{4} $\delta$ 0), is asymmetric with respect to {\bf a} and {\bf b}. At present, we have no explanation as to the source of this asymmetry in a tetragonal compound.

Among Eu intermetallics of the type EuMX$_3$, where M is a $d$ metal and X Ge or Si, \eun\ is the only one where the magnetic structure, of spiral type, has been determined. We think that the germanides EuRhGe$_3$, EuIrGe$_3$, EuPtGe$_3$ and the silicide EuPtSi$_3$ \cite{euptsi3,euptge3,euirge3}, which show a low field dip in their magnetization curves, should also present a spiral magnetic structure. It is of interest to gather the information about the number of magnetic transitions and the magnetic structure of the low temperature phase in the EuMX$_3$ intermetallics (putative, except for \eun), as shown in Table \ref{mt}.
\begin{table}[ht]
\caption{\label{mt}Magnetic characteristics in the EuMX$_3$ series. A * denotes that the spiral plane is deduced from single crystal magnetization data, not from neutron diffraction measurements.}
\begin{tabular}{|c||c|c|} \hline
material                & spiral plane & nb. of transitions \\ \hline \hline
EuNiGe$_3$ (this work)    & (b,c),(a,c)  &        2           \\ \hline 
EuPtGe$_3$ \cite{euptge3} & (a,b)$^*$    &        1           \\ \hline
EuRhGe$_3$ \cite{euirge3} & (a,b)$^*$    &        1           \\ \hline
EuIrGe$_3$ \cite{euirge3} & (b,c),(a,c)$^*$  &    2           \\ \hline
EuPtSi$_3$ \cite{euptsi3} & (b,c),(a,c)$^*$  &    2           \\ \hline
EuRhSi$_3$ \cite{eutsi3}  & no spiral    &        2           \\ \hline
EuIrSi$_3$ \cite{eutsi3}  & ? (no single crystal) & 2           \\ \hline  
\end{tabular}
\end{table}
It comes out that all the studied EuMX$_3$ materials present a spiral structure, except EuRhSi$_3$, the situation in EuIrSi$_3$ being unknown since no single crystal could be grown. There seems to be a correlation between the number of transitions and the plane of the spiral structure: one observes one transition if the spiral lies in the (a,b) plane, and two transitions if the spiral lies in the (b,c) or (a,c) plane, or if there is no spiral. In all the compounds, the intermediate phase between the two transitions is an incommensurate modulated phase, probably collinear, as determined by M\"ossbauer spectroscopy. One can conjecture that a spiral lying in the (a,b) plane is more stable than a spiral in the (b,c) or (a,c) planes, which breaks the tetragonal symmetry, as mentioned above. In the latter case, the transition from paramagnetism would therefore occur first towards the intermediate phase, then to the spiral phase. 

\section{Conclusion}

We have studied the magnetic order in EuNiGe$_3$ versus temperature and magnetic field by single crystal neutron diffraction. Despite the strong Eu absorption and a limited dataset, the complete (B,T) phase diagram in the low temperature phase was extracted. The zero-field magnetic structure was found to be an equal moment helicoidal phase, with an incommensurate wave-vector {\bf k}=(\nicefrac{1}{4} $\delta$ 0), with $\delta \simeq 0.05$. Applying the field along the tetragonal axis, we found the peculiar behaviour that $\delta$ changes from 0.05 to 0.072 at 2\,T, where a first magnetization jump occurs, and vanishes at 3\,T, where the second magnetization jump takes place. All the structures were refined with good accuracy.

These results are in perfect agreement with previous macroscopic measurements (magnetization and magneto-resistivity). The local information extracted from neutron diffraction allowed us to identify an additional transition under magnetic field. Most of these features (except the small incommensurate component of the propagation vector) were well reproduced by a self-consistent mean field calculation.

\end{document}